# Reduction of friction by normal oscillations.
# I. Influence of contact stiffness.

M. Popov[1,2,3*], V. L. Popov[1,2,3], N. V. Popov[1]

[1]Berlin University of Technology, 10623 Berlin, Germany
[2]Tomsk Polytechnic University, 634050 Tomsk, Russia
[3]Tomsk State University, 634050 Tomsk, Russia
[*]*Corresponding author: m@popov.name*

The present paper is devoted to a theoretical analysis of sliding friction under the influence of oscillations perpendicular to the sliding plane. In contrast to previous works we analyze the influence of the stiffness of the tribological contact in detail and also consider the case of large oscillation amplitudes at which the contact is lost during a part of the oscillation period, so that the sample starts to "jump". It is shown that the macroscopic coefficient of friction is a function of only two dimensionless parameters – a dimensionless sliding velocity and dimensionless oscillation amplitude. This function in turn depends on the shape of the contacting bodies. In the present paper, analysis is carried out for two shapes: a flat cylindrical punch and a parabolic shape. Here we consider "stiff systems", where the contact stiffness is small compared with the stiffness of the system. The role of the system stiffness will be studied in more detail in a separate paper.

## 1. Introduction

The influence of vibration on friction is of profound practical importance [1]. This phenomenon is used in wire drawing [2], [3], press forming [4] and many other technological applications. Experimental studies of the influence of ultrasonic oscillations on friction started in the late 1950s [5]. In the subsequent years several illuminating works were performed using various techniques, e.g. measurement of electrical conductivity of the contact [6],[7]. Reduced friction has been observed both with oscillations in the contact plane (in-plane) [8] and perpendicular to it (out-of-plane) [9]. In the 2000s, interest in the interaction of friction and oscillations was promoted by applications such as traveling wave motors [10], [11] and the rapidly developing field of nanotribology [12], [13]. In recent years, detailed studies of the influence of ultrasonic oscillations and comparisons with various theoretical models have been performed by Chowdhury et al. [14] and by Popov et al. for in-plane oscillations [15] and by Teidelt et al. for out-of-plane oscillations [16]. The latter paper also includes a comprehensive overview of previous works in the field up to 2012.

The above works provided an empirical basis for a qualitative understanding of the influence of oscillations on friction. However, good quantitative correspondence between experimental results and theoretical models could never be achieved (see e.g. a detailed discussion in [30], so it is not clear whether we adequately understand the physics of this phenomenon. Even the question of which oscillation properties determine the reduction of friction force is still under discussion: While in the case of static friction it seems to be the amplitude of displacement oscillation [15], for sliding friction it is believed to be the amplitude of velocity oscillation [11]. In the following, we will show that, in general, friction under oscillation is determined by both of these parameters.

The main novelty of the present paper compared to earlier work on the influence of oscillation on friction is explicit consideration of the *contact stiffness*. The influence of the contact stiffness is closely related to a fundamental and still unresolved question about the physical nature of the characteristic length determining the crossover from static friction to sliding. In earlier works on this topic, it was assumed that this characteristic length is an intrinsic property of a frictional couple and that its physical nature is rooted in microscopic interactions between the surfaces [15]. However, later investigations suggested another interpretation. Studies of friction in stick-slip microdrives [17], [18] have shown that the static and dynamic behavior of drives can be completely understood and precisely described without any fitting parameters just by assuming that the characteristic length responsible for the "pre-slip" during tangential (in-plane) loading of a contact is equivalent to partial slip in a tangential contact of bodies with curved surfaces. This contact-mechanical approach was substantiated in [19] by a theoretical study of the influence of in-plane oscillations on the static force of friction. It was shown that the characteristic length is simply the indentation depth multiplied by the coefficient of friction. Later, it was found that this is valid independently of the shape of the contact and also holds true for rough surfaces [20]. This hypothesis of the purely contact mechanical nature of the pre-slip and of the characteristic amplitude was verified experimentally for a wide range of radii of curvature and applied forces in [21] and [22]. It was thus confirmed that describing friction under oscillation, including pre-slip, is basically a matter of correct contact mechanics and that the main governing parameter for both normal and tangential oscillation is the indentation depth. This realization also led to new generalizations in the physics of friction [23], [24], which, however, still need experimental verification.

In the present paper we utilize this new understanding of the importance of the precise contact mechanics and the key property of contact stiffness when considering the details of frictional processes. We focus our attention on the influence of normal (out-of-plane) oscillations on the macroscopic frictional force. We begin by looking at a simple system consisting of a single spring and a frictional point, then extend our analysis to the Hertzian (parabolic) contact using the Method of Dimensionality Reduction [25]. For simplicity we do not deal with system dynamics, and instead impose a forced oscillation of the indentation depth. This restricts our analysis to systems where the contact stiffness is small compared with the stiffness of the system as a whole and the inertia of the contact region thus does not play any role. An analysis involving system dynamics is published in the second part of this two-part paper.

Another contribution of this paper is the consideration of large oscillation amplitudes, when the indenter starts jumping. To our knowledge this case has not previously been considered in theoretical models.

## 2. Simplified one-spring model

Let us consider an elastic body that is brought into contact with a flat substrate and then subjected to a superposition of an oscillation in the direction normal to the substrate and movement with a constant velocity in the tangential direction. We will assume that Coulomb's law of friction with a constant coefficient of friction $\mu_0$ is valid in the contact. We first consider a very simple model consisting of a single spring with normal stiffness $k_z$ and tangential stiffness $k_x$. As the reference state, the unstressed state in the moment of first contact with the substrate is chosen. Let us denote the horizontal and vertical displacements of the upper point of the spring from the reference state by $u_x$ and $u_z$ and the horizontal displacement of the lower (contact) point by $u_{x,c}$. The upper point is forced to move according to

$$u_z = u_{z,0} - \Delta u_z \cos \omega t \quad \text{and} \quad \dot{u}_x = v_x \tag{1}$$

(see Fig. 1).

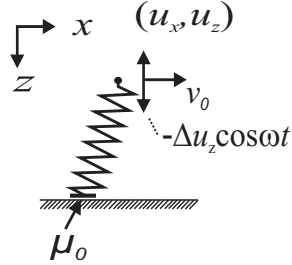

**Fig. 1** The simplest model of a tribological contact with a constant contact stiffness represented as a single spring, which has a normal stiffness $k_z$ and a tangential stiffness $k_x$. The upper end of the spring is forced to move according to (1). At the lower end (immediate contact spot), Coulomb's law of friction with a constant coefficient of friction $\mu_0$ is assumed.

### Small oscillation amplitudes (no "jumping")

Let us start our consideration with the case of sufficiently small oscillation amplitudes, $\Delta u_z < u_{z,0}$, so that the indenter remains in contact with the substrate at all times. As for the horizontal movement, the lower point of the spring can be either in stick or slip states. During the slip phase the tangential force $f_x = k_x(u_x - u_{x,c})$ is equal to the normal force $f_z = k_z(u_{z,0} - \Delta u_z \cos \omega t)$ multiplied with the coefficient of friction: $k_x(u_x - u_{x,c}) = \mu_0 k_z(u_{z,0} - \Delta u_z \cos \omega t)$. Differentiating this equation with respect to time gives $k_x(v_0 - \dot{u}_{x,c}) = \mu_0 k_z \omega \Delta u_z \sin \omega t$. For the tangential velocity of the lower contact point, it follows that $\dot{u}_{x,c} = v_0 - \mu_0 (k_z/k_x) \omega \Delta u_z \sin \omega t$. This equation is only valid when $\dot{u}_{x,c} > 0$, and the foot point of the spring will transition from the sliding state to the sticking state when the condition $\dot{u}_{x,c} = 0$ is fulfilled. This occurs at the time $t_1$ which satisfies the following equation:

$$\dot{u}_{x,c} = v_0 - \mu_0(k_z/k_x)\omega \Delta u_z \sin \omega t_1 = 0 \tag{2}$$

Introducing a dimensionless velocity

$$\bar{v} = \frac{k_x}{k_z} \frac{v_0}{\mu_0 \omega \Delta u_z}, \tag{3}$$

we can rewrite Eq. (2) in the form

$$\sin \omega t_1 = \bar{v}. \tag{4}$$

For $\bar{v} > 1$, this equation has no solutions, and the spring continues sliding at all times. Since, in this case, the tangential force remains proportional to the product of the normal force and the macroscopic coefficient of friction $\mu_0$ at all times, there is no reduction of the macroscopic force of friction.

For dimensionless velocities smaller than one, $\bar{v} < 1$, Eq. (4) has solutions and the movement of the contact point will consist of a sequence of sliding and sticking phases, where the sliding phase ends at time $t_1$ given by Eq. (4). The tangential force at this point is equal to $f_x = \mu_0 k_z(u_{z,0} - \Delta u_z \cos \omega t_1)$ or, taking (4) into account:

$$f_x(t_1) = \left(\mu_0 k_z u_{z,0} - \sqrt{(\mu_0 k_z \Delta u_{z,0})^2 - \left(\frac{v_x k_x}{\omega}\right)^2}\right). \tag{5}$$

During the sticking stage the tangential force increases linearly according to

$$f_x^{(stick)}(t) = f_x(t_1) + k_x v_x(t - t_1). \tag{6}$$

The next phase of slip starts at time $t_2$ when the tangential force becomes equal to the normal force multiplied by the coefficient of friction (see Fig. 2):

$$f_x(t_1) + k_x v_x(t_2 - t_1) = \mu_0 k_z(u_{z,0} - \Delta u_z \cos \omega t_2) \tag{7}$$

Or, taking (5) and (4) into account and using the dimensionless variable (3),

$$\cos \omega t_2 = \bar{v}(\tau_2 - \arcsin \bar{v}) + \sqrt{1 - \bar{v}^2}. \tag{8}$$

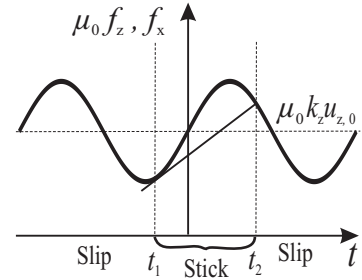

**Fig. 2** Schematic presentation of the normal force multiplied with the coefficient of friction (sinusoidal curve) and tangential force (straight line). During the slip phase (before $t_1$ and after $t_2$), the tangential force is equal to the normal force times the coefficient of friction, thus both curves coincide. During the stick phase (between $t_1$ and $t_2$), the tangential force is smaller than the normal force multiplied by the coefficient of friction. Both forces become equal again at time $t_2$, where the stick phase ends.

The average value of the frictional force during the whole oscillation period can be calculated as follows:

$$\langle f_x \rangle = \frac{\omega}{2\pi}\left[\int_{t_1}^{t_2} f_x^{(stick)}(t)dt + \int_{t_2}^{2\pi/\omega + t_1} f_x(t)dt\right]. \tag{9}$$

Divided by the average normal force, this gives the macroscopic coefficient of friction

$$\mu_{\text{macro}} = \langle f_x \rangle / \langle f_z \rangle, \tag{10}$$

where $\langle f_z \rangle = k_z u_{z,0}$ in the non-jumping case, which is considered here. The result of numerical evaluation of the macro-scopic coefficient of friction, normalized by the local coefficient of friction $\mu_0$ is presented in Fig. 3.

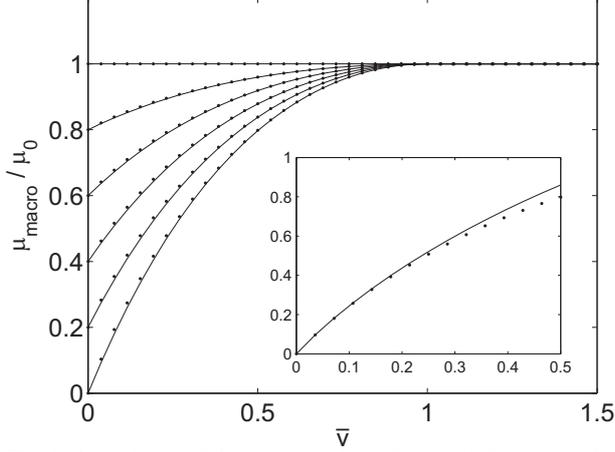

**Fig. 3** Dependence of the normalized coefficient of friction on the normalized velocity for $\Delta u_z / u_{z,0} = 0$; 0.2; 0.4; 0.6; 0.8; 1.0 (from top to bottom). Points represent the results of numerical evaluation of the integral (9). Solid lines represent the empirical approximation (11). The inset shows the low-velocity asymptotic solution (12) (solid line) compared to the numerical solution (9) (points).

It was found that the numerically obtained dependences of the coefficient of friction on dimensionless velocity and amplitude can be approximated very accurately with the following equation:

$$\frac{\mu_{\text{macro}}}{\mu_0} \approx 1 - \frac{\Delta u_z}{u_{z,0}} \left[ \frac{3}{4}(\bar{v}-1)^2 + \frac{1}{4}(\bar{v}-1)^4 \right]. \qquad (11)$$

A comparison of this approximation with numerical results provided by (9) and (10) is shown in Fig. 3. The low-velocity limit of the coefficient of friction can be derived analytically by replacing the time-dependence of the normal force with its Tailor series around the points $\omega t = 0$ and $\omega t = 3\pi/2$ and repeating the above calculations including integration of (9), which provides the result

$$\frac{\mu_{\text{macro}}}{\mu_0} = 1 + \frac{\Delta u_z}{u_{z,0}} \left( -1 + \pi \bar{v} - \frac{4}{3}\sqrt{\pi}\bar{v}^{3/2} + \frac{1}{2}\bar{v}^2 \right). \qquad (12)$$

This dependence is asymptotically exact in the limit of small sliding velocities. Like the empirical approximation (11) it contains only two dimensionless variables: the dimensionless amplitude of oscillation $\Delta u_z / u_{z,0}$ and the dimensionless sliding velocity (3). A comparison with the numerical results is shown for the case of the critical oscillation amplitude, $\Delta u_z / u_{z,0} = 1$, in the inset of Fig. 3.

Eq. (11) can be rewritten in a form explicitly giving the average tangential force (force of friction):

$$\langle f_x \rangle = \mu_0 k_z u_{z,0} \left( 1 - \frac{\Delta u_z}{u_{z,0}} \left[ \frac{3}{4}(\bar{v}-1)^2 + \frac{1}{4}(\bar{v}-1)^4 \right] \right). \qquad (13)$$

Note that the *change* of the friction force due to oscillation, as compared with sliding without oscillation, does depend on the amplitude of oscillation, but does *not* depend on the average normal force:

$$\Delta \langle f_x \rangle = -\mu_0 k_z \Delta u_z \left[ \frac{3}{4}(\bar{v}-1)^2 + \frac{1}{4}(\bar{v}-1)^4 \right]. \qquad (14)$$

As will be shown later, this property implies that Eq. (14) is valid for *arbitrarily-shaped* contacts if the oscillation amplitude is small and $k_x$ and $k_z$ are understood as the incremental tangential and normal stiffness of the contact.

Eq. (11) provides a compact representation of the law of friction. However, it is not always convenient for interpretation of experimental results, as the dimensionless velocity (3) is normalized by the amplitude of velocity oscillation and thus the scaling of the velocity depends on the oscillation amplitude. To facilitate the physical interpretation of experimental results it may be more convenient to normalize the velocity using a value that does not depend on the oscillation amplitude. Introducing a new normalized velocity $\hat{v}$ according to the definition

$$\hat{v} = \frac{k_x}{k_z} \frac{v_x}{\mu_0 \omega u_{z,0}}, \qquad (15)$$

we can rewrite (11) in the form

$$\frac{\mu_{\text{macro}}}{\mu_0} \approx 1 - \frac{\Delta u_z}{u_{z,0}} \left[ \frac{3}{4}\left(\hat{v}\frac{u_{z,0}}{\Delta u_z}-1\right)^2 + \frac{1}{4}\left(\hat{v}\frac{u_{z,0}}{\Delta u_z}-1\right)^4 \right]. \qquad (16)$$

This dependence is presented in Fig. 4.

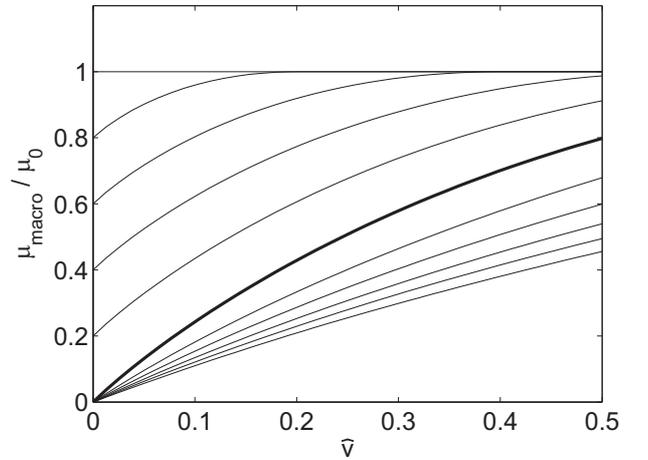

**Fig. 4** Dependence of the coefficient of the normalized friction on the dimensionless velocity (15): The horizontal line at the constant value 1 corresponds to sliding friction without oscillation. When the oscillation amplitude increases, the static force of friction (at zero velocity) decreases until it vanishes (bold line). This trend is shown in the upper part of the plot for amplitudes $\Delta u_z / u_{z,0} = 0.2$; 0.4; 0.6; 0.8; 1.0 (from top to bottom). Further increase of the oscillation amplitude leads to loss of contact during a part of the oscillation period. In this range of oscillation amplitudes, the static friction force remains zero, and the slope of the dependency decreases with increasing oscillation amplitude. This is shown in the lower part of the plot for amplitudes $\Delta u_z / u_{z,0} = 1.2$; 1.4; 1.6; 1.8; 2.0 (from top to bottom).

*Large oscillation amplitudes ("jumping")*

If the amplitude of normal oscillations exceeds the average indentation depth, the indenter starts to "jump": In this case it will be in contact with the substrate only during part of the oscillation period and in the no-contact-state for the rest of the time. For a jumping contact, analytical considerations become too cumbersome, and we will only present the results of numerical modeling of the behavior of the system. In our model, the movement of the contact point is determined by local stick and slip conditions: as long as the tangential spring force is smaller than the normal force multiplied by the coefficient of friction the contact point is considered to be stuck to the substrate. If in any particular time step the tangential force exceeds the maximum friction force it is brought into equilibrium by appropriately changing the contact coordinate. Overall, the system undergoes alternating contact and non-contact phases, while each contact phase may be divided into stick and slip phases. The average force during one complete period of oscillation, divided by the average normal force, results in the macroscopic coefficient of friction, $\mu_{macro}$. It can be shown that, as in the non-jumping case, the dimensionless coefficient of friction $\mu_{macro}/\mu_0$ is a function of only the dimensionless velocity $\bar{v}$ given by Eq. (3) and the dimensionless oscillation amplitude. This property was checked by varying (dimensional) system parameters while preserving the values of the two dimensionless parameters. The numerical results for the jumping case are shown in Fig. 5. One can see that the dependence of the reduced coefficient of friction on the reduced velocity does not change significantly after the reduced oscillation amplitude exceeds the critical value 1, where static friction first disappears. Thus, as a very rough approximation, one can use the relation (11) with the critical oscillation amplitude for the whole range of jumping contacts:

$$\frac{\mu_{macro}}{\mu_0} \approx 1 - \left[\frac{3}{4}(\bar{v}-1)^2 + \frac{1}{4}(\bar{v}-1)^4\right] \text{ (for the jumping case).} \quad (17)$$

It is interesting to note that the critical value of the dimensionless velocity $\bar{v}$, after which there is continuous sliding and the macroscopic coefficient of friction coincides with the microscopic one, is equal to 1 both in the non-jumping and jumping regimes.

The low-velocity asymptote of the dependence of the coefficient of friction can be easily found analytically. It is instructive to do this for a better understanding of the details of the dependence and of possible deviations from the rough estimate (17). At sufficiently low velocities, the spring will stick as soon as it comes into contact with the substrate. From (1), we can see that the times when contact is lost or regained are determined by the equation

$$\omega t_{1,2} = \pm \arccos(u_0/\Delta u_z), \text{ for } |\Delta u_z| > u_0. \quad (18)$$

The spring comes into contact in fully relaxed state and is then moved with the constant velocity $v_0$ during the contact time $t_{contact} = 2\pi/\omega - 2t_2$. At low velocities the spring will remain in stick for almost the entire contact time, so that the average tangential force during the contact time can be estimated as $\langle F_x \rangle_{contact} = k_x v_0 t_{contact}/2$ and the average tangential force during the whole oscillation period as

$$\langle F_x \rangle = \frac{k_x v_0 \omega t_{contact}^2}{2 \cdot 2\pi} = \frac{k_x v_0}{\pi \omega}\left(\pi - \arccos\left(\frac{u_0}{\Delta u_z}\right)\right)^2. \quad (19)$$

The average normal force is given by:

$$\langle F_z \rangle = k_z u_0 \left(1 - \frac{1}{\pi}\arccos\left(\frac{u_0}{\Delta u}\right) + \frac{1}{\pi}\frac{\Delta u}{u_0}\sqrt{1-\left(\frac{u_0}{\Delta u}\right)^2}\right) \quad (20)$$

with which we finally find the normalized coefficient of friction:

$$\frac{\mu_{macro}}{\mu_0} = \bar{v}\frac{\left(1-\frac{1}{\pi}\arccos\left(\frac{u_0}{\Delta u_z}\right)\right)^2}{\frac{u_0}{\Delta u_z}\left(1-\frac{1}{\pi}\arccos\left(\frac{u_0}{\Delta u_z}\right) + \frac{1}{\pi}\frac{\Delta u}{u_0}\sqrt{1-\left(\frac{u_0}{\Delta u_z}\right)^2}\right)}. \quad (21)$$

This result illustrates once more that the reduced coefficient of friction is a function of only the dimensionless velocity $\bar{v}$ and the dimensionless oscillation amplitude $\Delta u_z/u_{z,0}$. The dependence of the slope of the low-velocity asymptote on the dimensionless oscillation amplitude $\Delta u_z/u_{z,0}$ is shown in the inset of Fig. 5. On can see that when the sample starts jumping the slope drops rapidly by about 20% and then it remains nearly constant at the limiting value of $\pi/4$, thus an explicit expression for the low-velocity asymptote in the jumping regime can be written (in the original dimensional variables) as:

$$\mu_{macro} \approx \frac{\pi}{4}\frac{k_x}{k_z}\frac{v_0}{\omega \Delta u_z} \text{ (low velocity asymptote; } \Delta u_z > u_{z,0}\text{).} \quad (22)$$

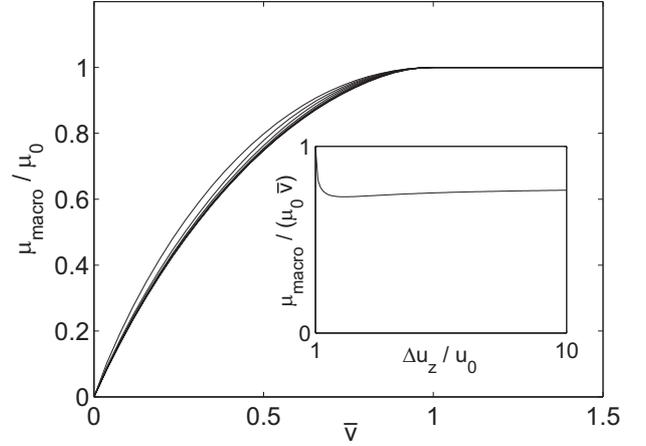

**Fig. 5** Dependence of the normalized coefficient of friction on the dimensionless velocity $\bar{v}$ (3) for the "jumping" case, i.e. when the oscillation amplitude exceeds the average indentation depth. Curves are shown for 11 oscillation amplitudes from $\Delta u_z/u_{z,0}=1$ to $\Delta u_z/u_{z,0}=2$ with a step of $0.1$. The curves for higher oscillation amplitudes "pile up" towards a limiting curve. The inset shows the dependence of the slope of the low-velocity asymptote (21) on $\Delta u_z/u_{z,0}$. One can see that it depends only weakly on the oscillation amplitude: Once the sample starts jumping the slope drops rapidly by about 20% and then remains practically constant with a limiting value of $\pi/4$.

As mentioned above, for comparison with experiments it may be preferable to use the dimensionless velocity $\hat{v}$ (15), which does not depend on the oscillation amplitude. In terms of this velocity, the coefficient of friction is shown in Fig. 4 for both jumping and non-jumping regimes, separated by a bold solid line corresponding to the critical amplitude $\Delta u_z/u_{z,0}=1$.

Overall, one can see that an increase of the oscillation amplitude first leads to a decrease of the static coefficient of friction at low sliding velocities. At the critical amplitude, the static coefficient of friction vanishes and remains zero during further increases of the oscillation amplitude, while the overall dependence on velocity starts to "tilt" (the slope of the dependence decreases with increasing oscillation amplitude).

## 3. Reduction of friction in a Hertzian contact

In Section 2, we considered a simplified model in which it was assumed that the contacting bodies have a constant contact stiffness that does not depend on the indentation depth. This model can be realized experimentally by using a flat-ended cylindrical pin or a curved body with a flat end (e.g. due to wear). However, in the general case the body in contact will have curved or rough surfaces so that the contact stiffness will depend on the indentation depth. In this section we generalize the results obtained in the previous section for more general contact configurations.

In our analysis of the contact of a curved body with the substrate we will use the so-called Method of Dimensionality Reduction (MDR). As shown in [26], the contact of arbitrarily shaped bodies can be described (in the usual half-space approximation ) by replacing it with a contact of an elastic foun-dation with a properly defined planar shape $g(x)$, as shown in Fig. 6. The elastic foundation consists of a linear arrangement of independent springs with normal stiffness $\Delta k_z$ and tangential stiffness $\Delta k_x$ and with sufficiently small spacing $\Delta x$. For an exact mapping, the stiffness of the springs has to be chosen according to [25], [27]:

$$\Delta k_z = E^* \Delta x \quad \text{with} \quad \frac{1}{E^*} = \frac{1-\nu_1^2}{E_1} + \frac{1-\nu_2^2}{E_2}, \quad (23)$$

$$\Delta k_x = G^* \Delta x \quad \text{with} \quad \frac{1}{G^*} = \frac{2-\nu_1}{4G_1} + \frac{2-\nu_2}{4G_2}, \quad (24)$$

where $E_1$ and $E_2$ are the moduli of elasticity, $\nu_1$ and $\nu_2$ the Poisson numbers, and $G_1$ and $G_2$ the shear moduli of the bodies. The "equivalent shape" $g(x)$ providing the exact mapping can be determined either analytically (e.g. for axis-symmetric profiles, [25], [27]), or by asymptotic [26], numerical [26],[28] or experimental methods. It is important to note that an equivalent profile does exist for arbitrary topographies of contacting bodies. Once determined, this equivalent profile can be used to analyze arbitrary dynamic normal and tangential loading histories. If the body is moved tangentially, the same law of friction that is valid for the three-dimensional bodies is applied for each individual spring, using the same coefficient of friction $\mu_0$. If the above rules are observed, the relations between macroscopic properties of the contact (in particular the normal and tangential force-displacement-relationships) will identically coincide with those of the initial three-dimensional problem [26].

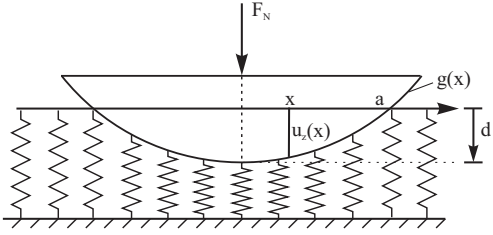

**Fig. 6** Schematic presentation of the contact of a transformed planar profile with an effective elastic foundation as prescribed by the rules of the Method of Dimensionality Reduction (MDR).

### Arbitrary surface topography and small amplitude of oscillations

Let us start by deriving the reduction of friction force for the case of arbitrary contact geometry and *small* oscillation amplitudes. Consider the MDR-representation of the problem in Fig. 6. If the oscillation amplitude is small, then most springs which came into contact with the elastic foundation during the initial indentation by $u_{z,0}$ will remain in contact at all times. Thus, the result (14), which is valid in the non-jumping case, is applicable for most of the springs in the contact; we only have to replace the normal contact stiffness by the stiffness of a single spring:

$$\Delta \langle f_x \rangle_{\text{one spring}} = -\mu_0 \Delta k_z \Delta u_z \left[ \frac{3}{4}(\bar{v}-1)^2 + \frac{1}{4}(\bar{v}-1)^4 \right]. \quad (25)$$

The oscillation amplitude and the expression in the brackets are the same for all springs. Summing over all springs therefore just means replacing the stiffness of one spring by the total stiffness of all springs in contact, $k_z$ which leads us back to Eq. (14), which is thus generally valid for arbitrary contact shapes.

### Parabolic surface profile and arbitrary amplitude of oscillations

For a parabolic profile $z = r^2/(2R)$ the equivalent plane profile $g(x)$ is given by [25]: $g(x) = x^2/R$. In our numerical simulations, this profile was first indented by $u_{z,0}$. Subsequently, the indenter was subjected to superimposed normal oscillation and tangential movement with constant velocity according to (1). Since the springs of the MDR model are independent, the simulation procedure for each spring is exactly as described in Section 2: The movement of the contact point of each spring of the elastic foundation was determined by the stick and slip conditions: as long as the tangential spring force remained smaller than the normal force multiplied by the coefficient of friction, the contact point remained stuck to the substrate. If in a particular time step the tangential force exceeded the critical value it was reset to the critical value by appropriately changing the contact coordinate. This procedure unambiguously determines the normal and tangential force in each spring of the elastic foundation at each time step. By summing the forces of all springs the total normal and tangential force are determined. After averaging over one period of oscillation, the macroscopic coefficient of friction was found by dividing the mean tangential force by the mean normal force. This coefficient of friction, normalized by the local coefficient of friction $\mu_0$, once again appears to be a function of only two parameters: the dimensionless velocity (either $\bar{v}$ (3), see Fig. 7, or $\hat{v}$ (15), see Fig. 8) and the dimensionless oscillation amplitude $\Delta u_z / u_{z,0}$.

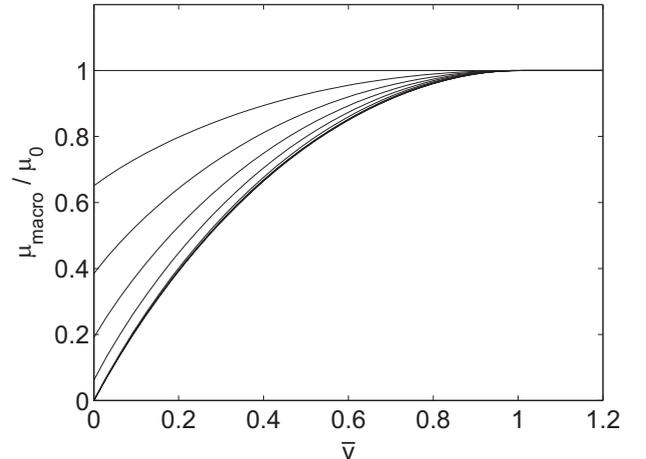

**Fig. 7** Dependence of the normalized coefficient of friction on the normalized velocity $\bar{v}$ defined by (3) for the oscillation amplitudes $\Delta u_z / u_{z,0} = 0$; 0.2; 0.4; 0.6; 0.8; 1.0 (top to bottom).

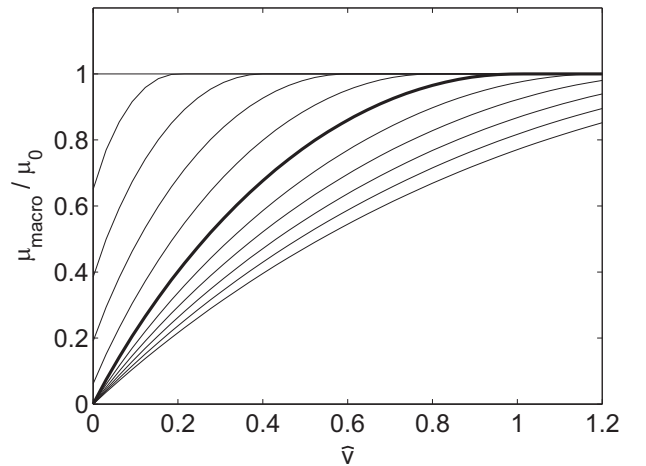

**Fig. 8** Dependence of the normalized coefficient of friction on the normalized velocity $\hat{v}$ defined by (15) for the oscillation amplitudes $\Delta u_z / u_{z,0} = 0$; 0.2; 0.4; 0.6; 0.8; 1.0 (the bold line and all curves in the upper-left part) and for $\Delta u_z / u_{z,0} = 1.2$; 1.4; 1.6; 1.8; 2.0 (bottom-right part).

For a parabolic profile, the dependences look qualitatively very similar to those for a single spring (compare with Fig. 3 and Fig. 4.) The dependences have two characteristic features: (a) the static force of friction – the starting point of the curve at zero velocity and (b) the critical velocity $\overline{v} = 1$ after which there is no further influence of oscillations on the macroscopic coefficient of friction.

## 4. Discussion

Let us summarize and discuss the main findings of the present study and provide a comparison with experimental results. The structure of the obtained dependences of the macroscopic coefficient of friction on the velocity in the presence of oscillations is simple and contains only two main reference points: The static friction force and the critical velocity. The dependence of the static friction force is extremely simple: it is determined just by the minimum of the normal force during the oscillation cycle. The differences of the static friction force for indenters of different shape will therefore be completely determined by the solution of the corresponding normal contact problem. The second reference point is the critical velocity, which separates the velocity interval where the coefficient of friction does depend on the velocity from the interval where there is no further dependence. This critical point is given by the condition $\overline{v} = 1$ or explicitly, in dimensional variables:

$$v_0 = \frac{E^*}{G^*} \mu_0 \omega \Delta u_z . \qquad (26)$$

Since Mindlin's ratio $E^*/G^*$ is on the order of unity and $\omega \Delta u_z$ is the amplitude of *velocity oscillation*, this means that the critical velocity is roughly speaking the amplitude of the velocity oscillation multiplied with the coefficient of friction. It is astonishing that this simple result is absolutely universal and is valid for both in the non-jumping and jumping regimes and for indenters of arbitrary shape.

Thus, one of the reference points is determined solely by the amplitude of displacement oscillation and the other solely by the amplitude of the velocity oscillation. Between these points, the dependence of the coefficient of friction on sliding velocity is accurately approximated by Eq. (11), which can be rewritten in a universal form that does not depend on the indenter shape:

$$\frac{\mu_{\text{macro}}}{\mu_0} \approx 1 - \left(1 - \frac{\mu_{\text{static}}}{\mu_0}\right)\left[\frac{3}{4}(\overline{v}-1)^2 + \frac{1}{4}(\overline{v}-1)^4\right]. \qquad (27)$$

The indenter shapes will only influence the static coefficient of friction in the above equation.

For practical applications one can use an even simpler approximation differing from (27) by 1% or less:

$$\frac{\mu_{\text{macro}}}{\mu_0} \approx 1 - \left(1 - \frac{\mu_{\text{static}}}{\mu_0}\right)(1-\overline{v})^{2.4} . \qquad (28)$$

Substituting the definition of $\overline{v}$, we can write this dependence in the initial dimensional variables:

$$\mu_{\text{macro}} \approx \mu_0 - (\mu_0 - \mu_{\text{static}})\left(1 - \frac{G^*}{E^*}\frac{v_0}{\mu_0 \omega \Delta u_z}\right)^{2.4} . \qquad (29)$$

This equation contains in a condensed form all essential results of the present study. Most interestingly, it is approximately valid in both non-jumping and jumping regimes and for all indenter shapes. As long as the amplitude of oscillation is smaller than the average indentation (no jumping), the static friction force decreases monotonously with increasing oscillation amplitude. After reaching the critical oscillation amplitude, the static friction force vanishes and remains zero at larger oscillation amplitudes, but Eq. (27) still remains valid in a good approximation. From the critical amplitude onwards, the force-velocity dependencies start to "tilt".

The described features can be readily recognized in the experimental data shown in Fig. 9.

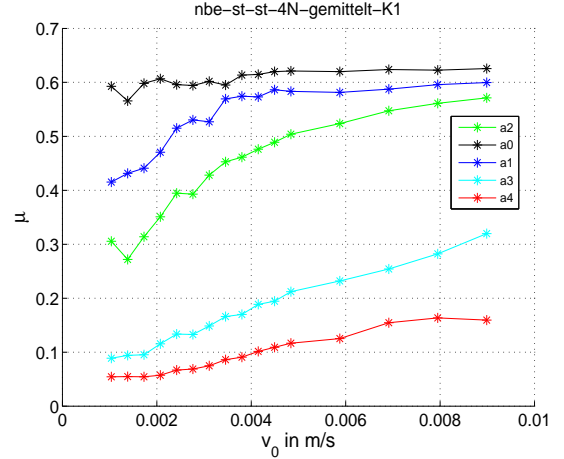

**Fig. 9** Experimentally determined dependences of the coefficient of friction on sliding velocity between a steel sphere and a steel disc on for increasing amplitudes of out-of-plane oscillation obtained by Milahin (Source: [29], reproduced with permission of the author). The upper-most curve corresponds to sliding in the absence of oscillation. The second, third and fourth curves correspond to amplitudes of $\Delta u_z = 0.06\ \mu\text{m}$, $\Delta u_z = 0.10\ \mu\text{m}$, $\Delta u_z = 0.16\ \mu\text{m}$ and $\Delta u_z = 0.27\ \mu\text{m}$.

Comparison of the experimental results with the theoretical predictions in Fig. 4 shows both similarities and differences. For example, we also observe the decrease of static friction and subsequent "tilting" of the dependences in the experimental data. Similar behavior was also observed in [30]. A difference between our theory and experiment is that the static coefficient of friction does not vanish entirely even at large oscillation amplitudes. This effect is known also for other modes of oscillation and is related to the microscopic heterogeneity of the frictional system, which means that Coulomb's law of friction is not applicable at very small space scales [31].

As we noted in the introduction, we considered a case of a soft contact and a rigid measuring system. In the opposite case of a very stiff contact and soft surrounding system, the dependences of the coefficient of friction on the oscillation amplitude appear to be essentially influenced by the inertial properties of the system [16]. An analysis carried out by Teidelt in [30] has shown that for the measuring system described in [16] a reasonable agreement between experiment and theory can only be achieved if the contact stiffness is taken into account. However, under other conditions – and in particular depending on the frequency of oscillations – the assumption of soft contact can fail. For such cases, a more general analysis has to be carried out, which will be provided in a separate paper.

## 5. Acknowledgements

This work was supported in part by the Ministry of Education of the Russian Federation, by COST Action MP1303 and the Deutsche Forschungsgemeinschaft.